\documentclass[pr, preprintnumbers, showpacs,longbibliography,footnoteadded,nofootinbib,]{revtex4-1}
\pdfoutput=1

\usepackage[T1]{fontenc}
\usepackage{amssymb}
\usepackage{graphicx}
\usepackage{amsmath}
\usepackage[
      colorlinks=true,
      linkcolor=blue,
      urlcolor=blue,
      filecolor=black,
      citecolor=blue,
            ]{hyperref}
\usepackage{tensor}
\usepackage{subfigure}
\usepackage{epstopdf}
\usepackage{color}

\usepackage{paralist}

\newcommand{\td}{\text{d}}

\newtheorem{theorem}{Theorem}

\begin{document}
\title{Constraining the number of horizons with energy conditions}

\author{Run-Qiu Yang }
\email{aqiu@tju.edu.cn (corresponding author)}
\affiliation{Center for Joint Quantum Studies and Department of Physics, School of Science, Tianjin University, Yaguan Road 135, Jinnan District, 300350 Tianjin, China}

\author{Rong-Gen Cai}
\email{cairg@itp.ac.cn}
\affiliation{CAS Key Laboratory of Theoretical Physics, Institute of Theoretical Physics,
Chinese Academy of Sciences, Beijing 100190, China}
\affiliation{School of Fundamental Physics and Mathematical Sciences, Hangzhou Institute for Advanced Study, UCAS, Hangzhou 310024, China}

\author{Li Li}
\email{liliphy@itp.ac.cn (corresponding author)}
\affiliation{CAS Key Laboratory of Theoretical Physics, Institute of Theoretical Physics,
Chinese Academy of Sciences, Beijing 100190, China}
\affiliation{School of Fundamental Physics and Mathematical Sciences, Hangzhou Institute for Advanced Study, UCAS, Hangzhou 310024, China}

\begin{abstract}
We show that the number of horizons of static black holes can be strongly constrained by energy conditions of  matter fields. After a careful clarification on the ``interior'' of a black hole, we prove that if the interior of a static  black hole satisfies strong energy condition or null energy condition, there is at most one non-degenerated inner Killing horizon behind the non-degenerated event horizon. Our result offers some universal restrictions on the number of horizons. Interestingly and importantly, it also suggests that  matter not only promotes the formation of event horizon but also prevents the appearance of multiple horizons inside black holes. Furthermore, using the geometrical construction, we obtain a radially conserved quantity which is valid for general static spacetimes.
\end{abstract}
\maketitle
\tableofcontents



\section{Introduction}
Black hole is one of the most fascinating objects predicted by general relativity and has attracted great interests in both theoretical and observational aspects.  The recent observations of gravitational waves by advanced LIGO~\cite{Abbott:2016blz} and the photos of the shadow of a black hole~\cite{Akiyama:2019cqa,Akiyama:2019brx,Akiyama:2019bqs} give us the definite evidence of the existence of black holes as well as offer us new and powerful venues to study black holes.  While the exterior of black holes has been widely investigated and many important properties were established (see \emph{e.g.} Refs.~\cite{Wald:1999vt,Frolov2015} for reviews), the structure behind the event horizon is lack of enough study and still mysterious even in theory. Definitely, the inner structure of black hole  is an interesting problem in its own right  and is important for black hole physics, gravitation and quantum physics. This has been strengthened by recent developments. For example, the interior of black hole is crucial in holographic computational complexity~\cite{Susskind:2014rva,Alishahiha:2015rta,Brown:2015bva} and recent proposals toward the resolution of information loss paradox~\cite{Christodoulou:2014yia,Bengtsson:2015zda,Almheiri:2019hni,Penington:2019npb,Almheiri:2019qdq}.


When we consider the interior of a black hole, a basic question arises: how many different horizons can appear in a black hole?  It is well known that the Schwarzschild black hole has only one horizon (event horizon). The Kerr-Newman  (KN) black hole has at most two different classes of horizons, \emph{i.e.}, an event horizon and an inner horizon (Cauchy horizon).  In fact, it is not difficult to find solutions with multiple horizons in general relativity. For example, one can construct multiple extreme Reissner-Nordstr\"{o}m (RN) black holes~\cite{Majumdar1947,Hartle1972}. In asymptotically de Sitter (dS) spacetime, the non-extreme  Kerr-Newman-dS black hole has three horizons and one can also construct black hole solutions with more horizons~\cite{Kastor1993,Brill1994,Nakao1995}. For more examples of black hole with multiple horizons, one can see Refs.~\cite{Bronnikov:2012mf,Gao:2017vqv,Nojiri:2017kex} and references therein. From the mathematical viewpoint, it is not surprising that one can construct a black hole with arbitrary number of horizons~\footnote{For particular forms of Lagrangian of matter, one can prove that there is no inner horizon for static black holes, see \emph{e.g.},~\cite{Hartnoll:2020rwq,Hartnoll:2020fhc,Cai:2020wrp,Devecioglu:2021xug,An:2021lmf}.}: one  starts with writing down a desired metric directly and then computes the corresponding energy-momentum tensor.  However,  the solutions obtained in this way may not describe any physically realizable black hole. It is widely believed that at least in the classical level the matter should satisfy some constraints named ``energy conditions'',  which crudely describe properties common to all (or almost all) states of matter that are well-established in physics but are sufficiently strong to rule out many unphysical ``solutions'' of the Einstein's equation. For example, the original singularity theorems of Penrose~\cite{Penrose:1964wq} and Hawking~\cite{Hawking:1966sx} were proved for matter obeying the null energy condition (NEC) or strong energy condition (SEC), respectively. Interestingly and surprisingly,  in this work we are able to show that the number of different horizons in a static black hole is also constrained strictly by SEC or NEC.

By definition, a horizon in static spacetime is a smooth null codimensional-1 surface, which can be foliated by a series of spacelike codimensional-2 surfaces.  Such spacelike codimensional-2 surfaces are homeomorphic to each other and each of them is called a cross-section of the horizon. In the present study, we are mainly interested in black holes with non-degenerate Killing horizons, \emph{i.e.} their surface gravities are nonzero.\,\footnote{Black holes with degenerate Killing horizons will be discussed in Sec.~\ref{disc}.} Denote $\Gamma$ to be an arbitrary cross-section of black hole event horizon and $\xi^\mu$ to be the Killing vector which presents the static symmetry. We can prove the following theorem for a static $(d+1)$-dimensional black hole.
\begin{theorem}\label{theorem1}
If the Einstein's equation and one of the following three conditions are satisfied
\begin{enumerate}
\item[(C1)] $\Gamma$ is compact and  SEC is satisfied inside black hole;
\item[(C2)] $\Gamma$ is noncompact but the cross-section of black hole event horizon has hyperbolical or planar symmetry, and NEC is satisfied inside black hole;
\item[(C3)] $\Gamma$ is noncompact surface with nonpositive area-averaged scalar curvature, and NEC is satisfied inside black hole in the case of $d=3$;
\end{enumerate}
then there is at most one inner Killing horizon associated with $\xi^\mu$ inside every connected branch of black hole event horizon.
\end{theorem}
Here we define the ``area-averaged scalar curvature'' of a surface $S$ as $ \mathcal{A}^{-1}\int_S \mathfrak{R}\td S$ with $ \mathcal{A}$ and $\mathfrak{R}$ the area and scalar curvature of $S$, respectively. When the cross-section is not compact, this integration should be defined by a suitable limit, which will be explained in Sec.~\ref{folia1}. The precise meaning of the ``interior'' will be given in Sec.~\ref{sec2}.

Our results offer universal constraints on the number of horizons inside a black hole independent of the details of matter. In particular, we are able to constrain the number of horizons without symmetry assumptions for C1 and C3. We also clarify a fact that has not been noticed for a long time. Differing from the naive stereotype that classical matter promotes the formation of horizons, our theorem shows that classical matter in fact plays two opposite roles: on the one hand, it triggers the formation of a black hole and thus leads to the appearance of horizon; on the other hand, it also prevents the formation of multiple horizons inside. In what follows we will first clarify the conceptions and assumptions needed for the precise statement of \textbf{Theorem}~\ref{theorem1} and then give the basic idea of the proof.

\section{Conception and assumption}\label{sec2}
{We first assume that the event horizon is a Killing horizon associated to the Killing vector $\xi^\mu$. In fact, it has been shown that, in wide variety of cases of interest, a black hole event horizon in the static case is the Killing horizon associated to $\xi^\mu$~\cite{Wald:1999vt}.} To determine the number of Killing horizons associated with $\xi^\mu$, one may study how many foots of $\xi^2=\xi^\mu\xi_\mu$ may appear. The question proposed in this way may lead to ambiguity.\,\footnote{More discussions are presented in appendix~\ref{rncase}.}
\begin{figure}
 \includegraphics[width=0.15\textwidth]{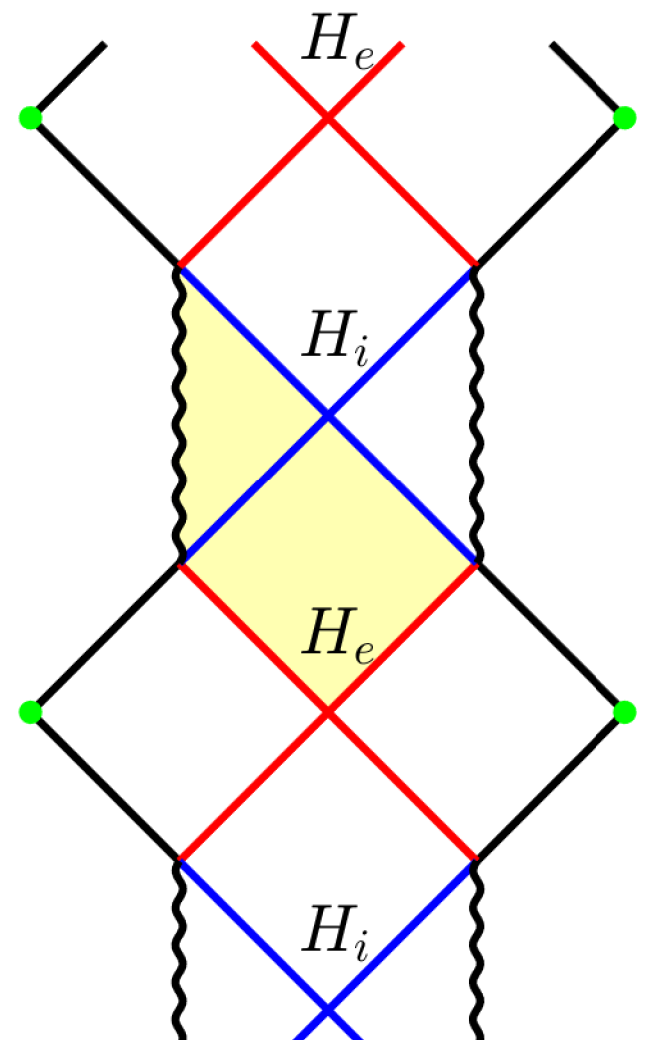}
   \caption{Penrose diagram of the maximally continued RN black hole solution. There are infinitely many disconnected event horizons ($H_e$) and inner horizons (Cauchy horizons) ($H_i$). The yellow region is called  the ``interior''  of a black hole in this work. }\label{figrn1}
\end{figure}
Before proceeding further, it is necessary to give a definite specification on the meaning of the number of horizons in the present work.

Instead of considering a specific coordinate system, we take advantage of the maximally analytically continued spacetime. Nevertheless, as can be seen, for example, from the Penrose diagram of RN metric in Fig.~\ref{figrn1}, there may exist an infinite lattice of universes connected by black hole tunnels and each universe contains its event horizon and Cauchy horizon. To avoid such trivial infiniteness due to analytical continuation, we count the number of horizons ``\textit{inside a connected branch of black hole event horizon}'' for which we clarify as follows.\,\footnote{One should not confuse the ``connected/disconnected'' and ``causally connected/disconnected''. The former is in the topological sense, while the later involves the causal structure of the spacetime. In this paper, when refer to  ``connected/disconnected'', we always mean the topological sense.}
Assume that $H_e^+$ is one connected branch of future event horizon $H_e$ (if $H_e^+$ is separated into two branches by a bifurcated surface, we only consider one of two branches).  In this paper, a region is called to be the inside  of $H_e^+$ if it is covered by future-directed infalling null geodesics starting from $H_e^+$ (see, for example, the yellow region of Fig.~\ref{figrn1}). In contrast to the timelike geodesics which in many cases could go from one universe to another by passing through horizons many times, the infalling null geodesics converge towards the singularity and thus are not able to connect to other black hole tunnels. Using this definition, one immediately obtains that the Schwarzschild black hole has one horizon, while the KN and RN black holes have two horizons, precisely agreeing with the usual statement  in the literature.

There is also another reason why we do not use timelike geodesics to define the ``inside/interior''. Let us consider a $(3+1)$ dimensional static spacetime, of which the metric reads
\begin{equation}\label{checkg1}
  \td s^2=-U(r)\td t^2+\frac{\td r^2}{U(r)}+r^2(\td\theta^2+\sin^2\theta\td\phi^2)\,.
\end{equation}
Here $U(r)=0$ has three single-roots $\{r_3<r_2<r_1\}$, $U(r>r_1)>0$  and the spacetime has a spacelike singularity at $r=0$. Let's consider a geodesic observer who has jumped into such black hole and is now traveling in the universe with $r\in(r_3,r_2)$ where $U(r)>0$. Denoting the observer's four-velocity to be $V^\mu=\td x^\mu/\td\tau$ with $\tau$ the proper time, one obtains from $V^\mu V^\nu g_{\mu\nu}=-1$ that
%
\begin{equation}\label{uutime1}
  -U\left(\frac{\td t}{\td\tau}\right)^2+\left(\frac{\td r}{\td\tau}\right)^2/U\leq-1\,.
\end{equation}
One the other hand, the ``energy'' $E=-V^\mu\xi_\mu$ of this geodesic observer is a constant, and thus one has
\begin{equation}\label{uutime2}
  \left(\frac{\td r}{\td\tau}\right)^2\leq E^2-U\,.
\end{equation}
Taking $U_m$ to be the maximum in the region with $r\in[r_3,r_2]$, we see that the geodesic observer cannot pass through the third horizon if $E^2\leq U_m$. Without additional constraints, $U_m$ can be arbitrarily large. In order to pass through the third horizon (if it appears) in all cases, a geodesic observer with ``infinitely large $E$ is required. Timelike geodesics in this limit become null geodesics.


In the following discussion,  we will consider static spacetime (here ``static means that the spacetime metric is time-independent outside the event horizon of the black hole).  Unless otherwise stated, the ``horizon'' always means ``Killing horizon'' associated with $\xi^\mu$.  We assume that every connected branch of a black hole event horizon is simply connected. To avoid sinking into the mathematical sea too deeply, we also assume that the singularities, if exist, not only hide in the interior of a connected branch of the black hole event horizon but also hide inside the innermost horizon. Though this assumption is stronger than the usual ``weak cosmic censorship conjecture''~\cite{1969NCimR...1..252P,Hod2008,Gwak2021}, it is satisfied in most of known examples of black holes. Under this assumption, the spacetime will be regular between the inner horizons and event horizon.

We begin with the Einstein's equation in $(d+1)$-dimensional spacetime
\begin{equation}\label{Einstein}
G_{\mu\nu}=\hat{T}_{\mu\nu}\,,
\end{equation}
where we have used the convention $8\pi G_N=1$ and have also absorbed the cosmological constant term to the energy momentum tensor $\hat{T}_{\mu\nu}$. The SEC stipulates that
\begin{equation}\label{stronge1}
  [\hat{T}_{\mu\nu}-\hat{T}g_{\mu\nu}/(d-1)]v^\mu v^\nu\geq0\,,
\end{equation}
for every timelike vector field $v^\mu$, and the NEC says that $\hat{T}_{\mu\nu}k^\mu k^\nu\geq0$ for arbitrary null vector $k^\mu$. With the above clarifications, we are ready to prove \textbf{Theorem}~\ref{theorem1}.

\section{Proof of C1}
\begin{figure}
\centering
   \includegraphics[width=0.3\textwidth]{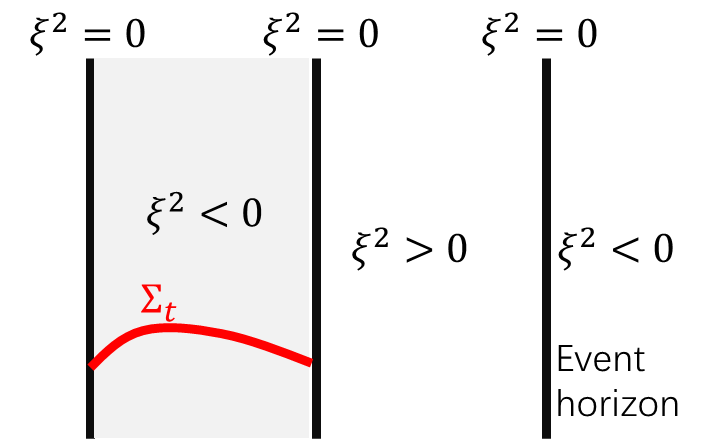}
   \caption{ Two non-degenerate inner Killing horizons appear inside a non-degenerate black hole event horizon. }\label{fighs1}
\end{figure}

Assume that the condition C1 is true but the conclusions of \textbf{Theorem}~\ref{theorem1} are wrong, then there must be  an connected spacetime region $\mathcal{V}$ inside the black hole event horizon such that the Killing vector $\xi^\mu$ is timelike inside $\mathcal{V}$ and is null at $\partial\mathcal{V}$ (See Fig.~\ref{fighs1} for example). We choose the orbit of $\xi^\mu$ as the time coordinate $t$ and denote $\Sigma_t$ to be equal-$t$ slice of $\mathcal{V}$, in which there is no any singularity.
The metric in region $\mathcal{V}$ has the following $(d+1)$ decomposition:
\begin{equation}\label{decomsigam1}
  \td s^2=-N^2\td t^2+h_{ab}\td x^a\td x^b\,,
\end{equation}
where $N$ and $h_{ab}$ are in general as functions of all coordinates except for $t$.
The Einstein's equation~\eqref{Einstein} in the static case contains the following two independent equations:
\begin{equation}\label{hamilt}
  {^{(d)}R}=2\hat{\rho}\,,
\end{equation}
and
\begin{equation}\label{eqdotK1}
  {^{(d)}R}_{ab}=N^{-1}D_aD_bN+\left[{\hat{\mathcal{T}}}_{ab}+\frac{h_{ab}}{d-1}(\hat{\rho}-\hat{\mathcal{T}})\right]\,.
\end{equation}
Here ${^{(d)}R}_{ab}$ and $D_a$ are the $d$-dimensional Ricci tensor and covariant derivative operator associated with $h_{ab}$, respectively. ${^{(d)}R}$ is the $d$-dimensional scalar curvature of $\Sigma_t$, $\hat{\rho}=N^{-2}\hat{T}_{\mu\nu}\xi^\mu\xi^\mu$ is the energy density, $\hat{\mathcal{T}}_{ab}$ is the projection of stress tensor $\hat{T}_{\mu\nu}$ at $\Sigma_t$ and $\hat{\mathcal{T}}$ is the trace of $\hat{\mathcal{T}}_{ab}$. These two equations come from the Hamiltonian constraint and evolutional equation of extrinsic curvature. Their combination gives
\begin{equation}\label{eqforphi2}
  D^2N=\frac{N}{d-1}[(d-2)\hat{\rho}+\hat{\mathcal{T}}]=\frac{N}{d-1}[(d-1)\hat{\rho}+\hat{T}]\,,
\end{equation}
which yields
\begin{equation}\label{eqforphi2b}
  D^2N^2=2N^2[\hat{\rho}+\hat{T}/(d-1)]+2h^{ab}(\partial_aN)(\partial_bN)\,.
\end{equation}
Here we have used $\hat{T}=-\hat{\rho}+\hat{\mathcal{T}}$. The SEC~\eqref{stronge1} implies $\hat{\rho}+\hat{T}/(d-1)\geq0$. Thus, inside $\Sigma_t$ we have
\begin{equation}\label{eqforphi3}
  D^2N^2\geq0\,.
\end{equation}
%
As the cross-section of horizons is compact, the domain $\Sigma_t$ is bounded. The maximum principle shows that the maximum of $N^2$ must be at the boundaries of $\Sigma_t$, so we have
\begin{equation}\label{maxN1}
  \max N^2|_{\Sigma_t}=\max N^2|_{\partial\Sigma_t}=0\,,
\end{equation}
which is contradictory to the fact that $\xi^\mu$ is timelike inside $\Sigma_t$. Thus  the conclusions of \textbf{Theorem}~\ref{theorem1} is true provided the condition C1 holds.

As an explicit example, let us consider the $(3+1)$ dimensional metric of Eq.~\eqref{checkg1}.
Since $r_2$ and $r_3$ are the second and third horizons with $r_2>r_3$ inside the event horizon, we have $U(r_2)=U(r_3)=0$ and $U(r)>0$ when $r\in(r_3,r_2)$. This means that $U(r)$ must have local maximum in the interval $(r_3,r_2)$. For the spacetime~\eqref{checkg1}, the SEC requires $\hat{p}_{\theta}\equiv {T_{\theta}}^\theta=U''/2+U'/r\geq0$ (note that here we are considering the four-dimensional case), implying that $U(r)$ cannot have local maximum in the interval $(r_3,r_2)$. Thus, the SEC must be violated somewhere between the second and third horizons for black holes with more than one inner horizon behind the event horizon. A proof for the spherically symmetric case can also be found in Ref.~\cite{Zaslavskii:2010qz}.


\section{Foliation on equal-$t$ hypersurface}\label{folia1}
\begin{figure}
\centering
    \includegraphics[width=0.35\textwidth]{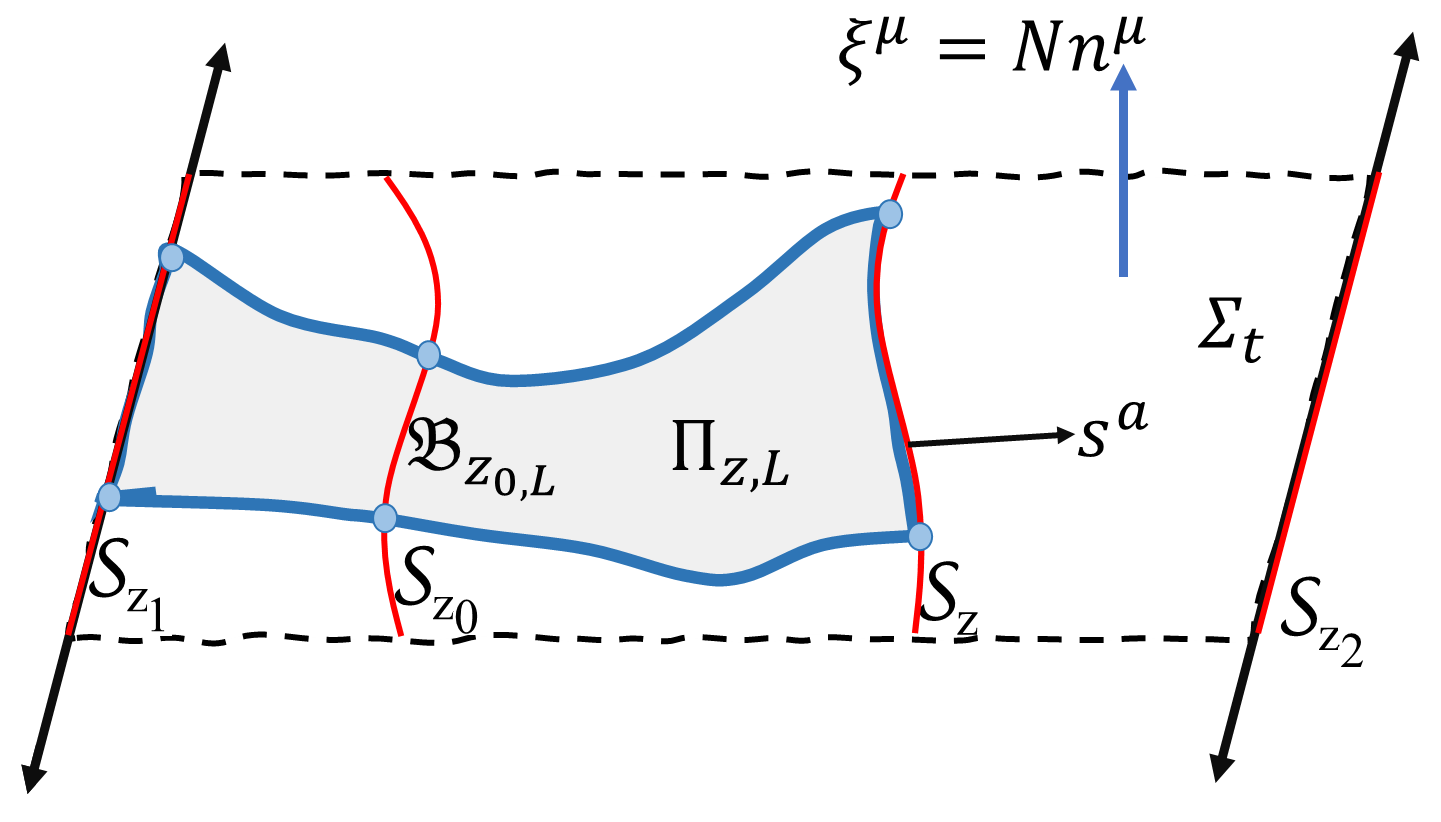}
   \caption{$\Sigma_t$ is foliated by a series of $(d-1)$-dimensional equal-$z$ surfaces $\{\mathcal{S}_z\}$(denoted by red lines). The shadow region is $\Pi_{z,L}=\cup_{z_0\in[z_1,z]}\mathfrak{B}_{z_0,L}$. }\label{figfoliation}
\end{figure}

To prove \textbf{Theorem}~\ref{theorem1} under the conditions C2 and C3, we need some additional preparations. We again assume that the conclusions of Theorem~\ref{theorem1} are wrong, then there must exist a spacelike equal-$t$ hypersurface $\Sigma_t$ bounded by two horizons. We consider the noncompact horizon case, so we can foliate $\Sigma_t$ by a series of non-compact $(d-1)$-dimensional spacelike surfaces $\{\mathcal{S}_z\}$ labeled by a radial coordinate $z$ (see Fig.~\ref{figfoliation}). Here $\mathcal{S}_z$ are all homeomorphic to $\mathbb{R}^{m}\times\mathbb{M}^{d-1-m}$ with $m$ a positive integer and $\mathbb{M}^{d-1-m}$ a closed $(d-1-m)$-dimensional space. Its bundary $\partial\Sigma_t$ locates at $z=z_1$ and $z=z_2>z_1$, respectively. Take $n^\mu$ to be the unit normal of $\Sigma_t$, $s^a$ to be unit normal of $\mathcal{S}_z$ and tangent to $\Sigma_t$, $\mathcal{K}_{AB}$  and $\mathfrak{R}$ to be extrinsic curvature and intrinsic scalar curvature of $\mathcal{S}_z$.
We require $s^a$ to be inward, \emph{i.e.}, $s^a\partial_az>0$ inside $\Sigma_t$. By introducing the local coordinates $x^a=\{z,x^A\}$, the induced metric of $\Sigma_t$ reads
\begin{equation}\label{induceqij1}
  \td s^2_{\Sigma_t}=\varphi^2\td z^2+q_{AB}(\td x^A+N^A\td z)(\td x^B+N^B\td z)\,,
\end{equation}
where $q_{AB}$ is the induced metric of $\mathcal{S}_z$, $\varphi$ is the lapse and $N_A$ is the shift vector.  Here $\varphi$ and $N^A$ are purely gauge freedoms and we are free to choose their values.

We define
\begin{equation}\label{defvarpi}
  \hat{p}_z:=\hat{\mathcal{T}}_{ab}s^a s^b, ~~\varpi:=\hat{\rho}d+\hat{T}-\hat{p}_z\,,
\end{equation}
and $\mathcal{K}=q^{AB}\mathcal{K}_{AB}$. Assuming that $\{x^1,x^2,\cdots,x^m\}$ are the noncompact coordinates, we take a compact $(d-1)$-dimensional subsurface $\mathfrak{B}_{z,L}:=\{(z,x^A)\in\mathcal{S}_z|-L\leq x^i\leq L,i=1,2,\cdots,m\}$ in every $\mathcal{S}_z$ and a domain $\Pi_{z,L}:=\cup_{z_0\in[z_1,z]}\mathfrak{B}_{z_0,L}$. Then we define
\begin{equation}\label{defAz1}
  \mathcal{A}(L):=z_1^{d-1}\int_{\mathfrak{B}_{z_1,L}}\td S\,,
\end{equation}
\begin{equation}\label{surfeq1}
  F(z):=\lim_{L\rightarrow\infty}\frac1{\mathcal{A}(L)}\int_{\partial\Pi_{z,L}-\mathfrak{B}_{z_1,L}}\left[\partial_aN-\frac{\mathcal{K}Ns_a}{d-1}\right]\td S^a,
\end{equation}
and
\begin{equation}\label{defintpi1}
\begin{split}
  I(z):=&\lim_{L\rightarrow\infty}\frac1{(d-1)\mathcal{A}(L)}\int_{\mathfrak{B}_{z,L}}\left[N\varphi(\varpi-\mathfrak{R})+\mathfrak{D}^2(N\varphi)-2q^{AB}(\partial_A\varphi)\partial_BN\right]\td S\,.
  \end{split}
\end{equation}
Here $\td S^a$ is the directed surface element, $\td S$ is the (scalar) surface element and $\mathfrak{D}^2$ the Laplace operator associated with $q_{AB}$. Note that the integration~\eqref{defAz1} is divergent when $L\rightarrow\infty$ due to the noncompactness of $\mathcal{S}_z$. However, the ratios~\eqref{surfeq1} and \eqref{defintpi1} are both well-defined. {The ``area-averaged scalar curvature'' mentioned in our \textbf{Theorem}~\ref{theorem1} is also understood in a similar limit.} Though $\partial\Pi_{z,L}-\mathfrak{B}_{z_1,L}\neq\mathfrak{B}_{z,L}$ for finite $L$, Eq.~\eqref{surfeq1} implies that in the limit $L\rightarrow\infty$ one has
\begin{equation}\label{surfeq2}
  F(z)=\lim_{L\rightarrow\infty}\frac1{\mathcal{A}(L)}\int_{\mathfrak{B}_{z,L}}\left[s^a\partial_aN-\frac{\mathcal{K}N}{d-1}\right]\td S\,.
\end{equation}
For simplicity, we will omit the notation ``$\lim_{L\rightarrow\infty}$'' and always assume $L\rightarrow\infty$ in what fallows.
For arbitrary lapse function $\varphi$ and shift  vector $N_A$,  the quantities $F(z)$ and $\varpi$ satisfy a few properties that are given as follows.

Firstly,  for an arbitrary point $q$, if NEC is satisfied at $q$, then
\begin{equation}\label{lemma1}
  \varpi|_{q}\geq0\,.
\end{equation}
We can prove Eq.~\eqref{lemma1} by following steps. Denoting the pull-back of the normal vector $s^a$ into spacetime to be $s^\mu$, we introduce $(d-1)$ orthogonal unit vectors $\{e^\mu_A\}$ and $\{n^\mu, s^\mu, e^\mu_A\}$ form a local Lorentz frame. By defining $\hat{p}_A=\hat{T}_{\mu\nu}e^{\mu}_Ae^\nu_A$, we have $\hat{T}=-\hat{\rho}+\hat{p}_z+\sum_A\hat{p}_A$.
As the spacetime is static, the energy momentum tensor satisfies $\hat{T}_{\mu\nu}n^\mu e^\nu_A=0$. Introducing $(d-1)$ different null vectors $l_A^\mu=n^\mu+e^\mu_A$, we find $\varpi=\sum_A(\hat{\rho}+\hat{p}_A)=\sum_A\hat{T}_{\mu\nu}l^\mu_Al^\nu_A\geq0$.

Secondly, if there are two inner horizons located at $z_1$ and $z_2$, i.e., $N|_{z_1}=N|_{z_2}=0$ and $N>0$ when $z\in(z_1,z_2)$, then we have $s^a\partial_aN|_{z=z_2}\leq0$ and so Eq.~\eqref{surfeq2} implies $F(z_2)\leq0$. Near the inner horizon $z=z_1$, we decompose the normal vector $s^a$ into $s^a=c(\partial/\partial z)^a+b^A(\partial/\partial x^A)^a$. The requirement $s^a\partial_az>0$ implies $c>0$. As $N>0$ when $0<(z-z_1)\ll 1$, we then have $N=N_m(z-z_1)^m+\cdots$ with $N_m>0$ and $m>0$. The leading term of Eq.~\eqref{surfeq2} reads
\begin{equation}\label{surfeq4}
  F(z)=(z-z_1)^{m-1}m\mathcal{A}^{-1}\int_{\mathcal{S}_{z}}cN_m\td S+\cdots>0\,.
\end{equation}
Thus, we obtain the second property
\begin{equation}\label{lemma2}
  \exists z_*\in(z_1,z_2)~~\text{such that}~F'(z_*)<0\,.
\end{equation}

Finally, in order to present the third property, let us define $\Delta\Pi_{z,L}:=\Pi_{z+\Delta z,L}-\Pi_{z,L}$ with $\Delta z\rightarrow0$ and so Eq.~\eqref{surfeq1} implies
\begin{equation}\label{surint1}
  F'(z)=\frac{F(z+\Delta z)-F(z)}{\Delta z}=\frac1{\mathcal{A}\Delta z}\int_{\partial\Delta\Pi_{z,L}}\left[\partial_aN-\frac{\mathcal{K}N s_a}{d-1}\right]\td S^a\,.
\end{equation}
With Gauss theorem we  can have
\begin{equation}\label{surint2}
\int_{\partial\Delta\Pi_{z,L}}\left[\partial_aN-\frac{\mathcal{K}Ns_a}{d-1}\right]\td S^a=\int_{\Delta\Pi_{z,L}}\left[D^2N-\frac{D_a(\mathcal{K}N s^a)}{d-1}\right]\td V\,,
\end{equation}
here $\td V$ denotes the volume element of $\Delta\Pi_{z,L}$. From the metric~\eqref{induceqij1} we find $\td V=\varphi\td S\td z$, where $\td S$ is the surface element of $\mathfrak{B}_{z,L}$.
On the other hand, we have
\begin{equation}\label{compkns2}
\begin{split}
  D^a(\mathcal{K}N s_{a})=Ns^a\partial_a \mathcal{K}+N\mathcal{K}^2+\mathcal{K}s^a\partial_a N\,,
  \end{split}
\end{equation}
and~\cite{Yoshino:2017gqv}
\begin{equation}\label{sdk1}
  s^a\partial_a \mathcal{K}=-\frac12({^{(d)}R}-\mathfrak{R}+\mathcal{K}^2+\mathcal{K}_{AB}\mathcal{K}^{AB})-\varphi^{-1}\mathfrak{D}^2\varphi\,,
\end{equation}
\begin{equation}\label{sdk2}
\mathcal{K}_{AB}\mathcal{K}^{AB}=\mathcal{K}^2-\mathfrak{R}-2G_{ab}s^a s^b+\frac2{N}\mathfrak{D}^2N+2\mathcal{K}N^{-1}s^a\partial_a N\,,
\end{equation}
where $G_{ab}$ is the projection of Einstein tensor on $\Sigma_t$. Putting Eq.~\eqref{sdk1} into Eq.~\eqref{compkns2} and using Eq.~\eqref{sdk2}, we obtain
%
\begin{equation}\label{surint4}
\begin{split}
&F'(z)=\frac1{(d-1)\mathcal{A}}\int_{\mathcal{S}_{z,L}}\left[(d-1)D^2N+{^{(d)}}RN/2\right.\\
&\left.-NG_{ab}s^as^b-\mathfrak{R}N+\varphi^{-1}N\mathfrak{D}^2\varphi+\mathfrak{D}^2N\right]\varphi\td S\,,
\end{split}
\end{equation}
where we have used $\td V=\varphi\td S\td z$. A concrete example about Eq.~\eqref{surint4} in the planar/hyperbolically/spherically symmetric horizon case can be found in appendix~\ref{app1}.  Note that  Eq.~\eqref{surint4} holds in any gravity theory. Focusing on Einstein's theory of general relatovity, we can use Eqs.~\eqref{hamilt}, \eqref{eqforphi2} and $G_{ab}s^as^b=\hat{\mathcal{T}}_{ab}s^as^b=\hat{p}_z$ to eliminate $D^2N, {^{(d)}}R$ and $G_{ab}s^as^b$ terms. Then we can obtain the following identity
\begin{equation}\label{fizeq1}
  F'(z)=I(z)\,.
\end{equation}

We stress that these three properties~\eqref{lemma1}, \eqref{lemma2} and \eqref{fizeq1} are also true when the event horizon is compact. Our Eq.~\eqref{fizeq1} implies $F(z)-\int^zI(y)\td y$ is dependent of the radial coordinate $z$. In the Einstein-Maxwell-complex scalar theory with spherical/planar/hyperbolical horizon symmetry, one can verify that $F(z)-\int^zI(y)\td y$ yields the  conserved charge proposed by Refs.~\cite{Hartnoll:2020fhc,Cai:2020wrp,Devecioglu:2021xug}. Our method is based on pure geometrical properties rather than the scaling symmetry which is used in Refs.~\cite{Hartnoll:2020fhc,Cai:2020wrp,Devecioglu:2021xug}. It is interesting to explore if there is any deep connection between our geometric construction and the scaling symmetry in the future.

\section{Proof of C2 and C3}
In the case of horizons with the planar or hyperbolical symmetry, we take  $N^2=z^{-2}f(z)e^{-\chi(z)}$ and foliate $\Sigma_t$ as follows
\begin{equation}\label{metric10}
  \mathrm{d}s_{\Sigma_t}^2=z^{-2}[f(z)^{-1}\td z^2+\td\Sigma_{k,d-1}^2]\,,
\end{equation}
with $\td\Sigma_{k,d-1}^2$ the metric for a planar $(k=0)$ or hyperbolic $(k=-1)$ surface.  Using the fact $\mathfrak{R}=(d-1)(d-2)kz^2$, we obtain
\begin{equation}\label{defQT1}
\begin{split}
  I(z)=\left[\frac{\varpi}{d-1}-(d-2)k z^2\right]z^{-1-d}e^{-\chi/2}\,.
  \end{split}
\end{equation}
Imposing the NEC, we have $\varpi\geq0$. We find that $F'(z)=I(z)\geq-(d-2)z^{1-d}e^{-\chi/2}k\geq0$ for all $z\in[z_1,z_2]$, which is contradictory to Eq.~\eqref{lemma2} if there is more than one inner horizon. This finishes our proof for the case C2.

%
We now prove the case C3 where the cross-section of event horizon $\Gamma$ is a noncompact 2-dimensional surface. Note that the properties~\eqref{lemma1}, \eqref{lemma2} and \eqref{fizeq1} are valid for arbitrary lapse function. We choose $\varphi=z^{-2}e^{-\chi(z)/2}N^{-1}$ with an arbitrary smooth function $\chi(z)$. Eq.~\eqref{defintpi1} then becomes
\begin{equation}\label{defintpi2a}
  I(z)=\frac1{(d-1)\mathcal{A}}\int_{\mathfrak{B}_{z,L}}[\varpi-z^2\mathfrak{R}+2q^{AB}N^{-2}(\partial_AN)\partial_BN]z^{-2}e^{-\chi/2}\td S\,.
\end{equation}
As $\varpi\geq0$ because of the NEC, one has $F'(z)\geq-e^{-\chi/2}\mathcal{A}^{-1}\int_{\mathfrak{B}_{z,L}}\mathfrak{R}\td S$. Both $\{\mathfrak{B}_{z,L}\}$ and $\Gamma$ are 2-dimensional surfaces and have the same topology, the uniformization theorem \cite{Forster1981} implies that they are conformal to each other with a conformal factor $e^{2\Omega(x^A,z)}$. The factor $e^{2\Omega(x^A,z)}$ measures the ratio of surface-area element between $\mathfrak{B}_{z,L}$ and $\Gamma$, which must be bounded. Otherwise the expansion of some null geodesics must be divergent between inner horizons and event horizon, which will lead to caustic singularity. We then find
\begin{equation}\label{scalarH1}
  \frac1{\mathcal{A}}\int_{\mathfrak{B}_{z,L}}\mathfrak{R}\td S=\frac1{\mathcal{A}}\int_{\Gamma}(\mathfrak{R}-\mathfrak{D}^2\Omega)\td S=\frac1{\mathcal{A}}\int_{\Gamma}\mathfrak{R}\td S\leq0\,.
\end{equation}
Here we used the assumption that ``area-averaged scalar curvature'' of $\Gamma$ is nonpositive. We also used the fact that the boundedness of $\Omega$ leads to $\mathcal{A}^{-1}\int_{\Gamma}\mathfrak{D}^2\Omega\td S\rightarrow0$~\footnote{When $\Omega$ is unbounded, this is not true. For example, the hyperbolic disc $\mathbb{D}^2$ is conformal to a flat plane $\mathbb{R}^2$ with metric $\td s_{\mathbb{D}^2}^2=z^{-2}\td s_{\mathbb{R}^2}^2$. Then we find $\mathfrak{D}^2\Omega=-2/z^2$ and so $\mathcal{A}_{\mathbb{D}^2}^{-1}\int_{\mathbb{R}^2}\mathfrak{D}^2\Omega\td S=-2\neq0$. Here $\mathcal{A}_{\mathbb{D}^2}$ is the area integration of ${\mathbb{D}^2}$. }.
Eq.~\eqref{scalarH1} implies $F'(z)\geq0$ for all $z\in[z_1,z_2]$, which is contradictory to Eq.~\eqref{lemma2}. Thus we end the proof for the case C3.



\section{Discussion}\label{disc}
In this paper, we have discussed the number of horizons for a static black hole. We proved that the number of horizons is strongly constrained by the SEC and NEC. We also found the inhomogeneous generalization of conserved charge proposed by Refs.~\cite{Hartnoll:2020fhc,Cai:2020wrp}. Note that our theorem does not require the corresponding energy conditions to be satisfied in the whole spacetime. In the asymptotically flat (3+1)-dimensional case, the Hawking's topology theorem~\cite{hawking1972,Hawking:1973uf,Galloway:2005mf} says that the cross-section of black hole horizon is sphere provided the dominant energy condition is satisfied. The planar or hyperbolic topology of black hole horizon may still appear if matter only satisfies SEC or NEC. For example, the so-called topological black hole with Ricci flat or hyperbolic horizon appears in anti-de Sitter (AdS) space.

%
\begin{figure}[h!]
\centering
    \subfigure[]{\includegraphics[width=0.3\textwidth]{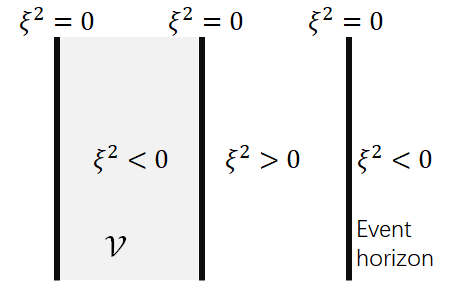}}
    \subfigure[]{\includegraphics[width=0.3\textwidth]{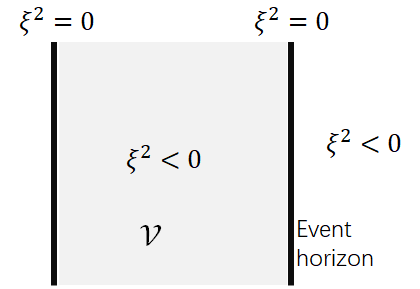}}
   \caption{(a) Configuration of two odd-order horizons that appear inside an odd-order event horizon; (b) Configuration of one inner horizon inside an even-order event horizon.  }\label{fighs1b}
\end{figure}

\begin{figure}
\centering
   \includegraphics[width=0.5\textwidth]{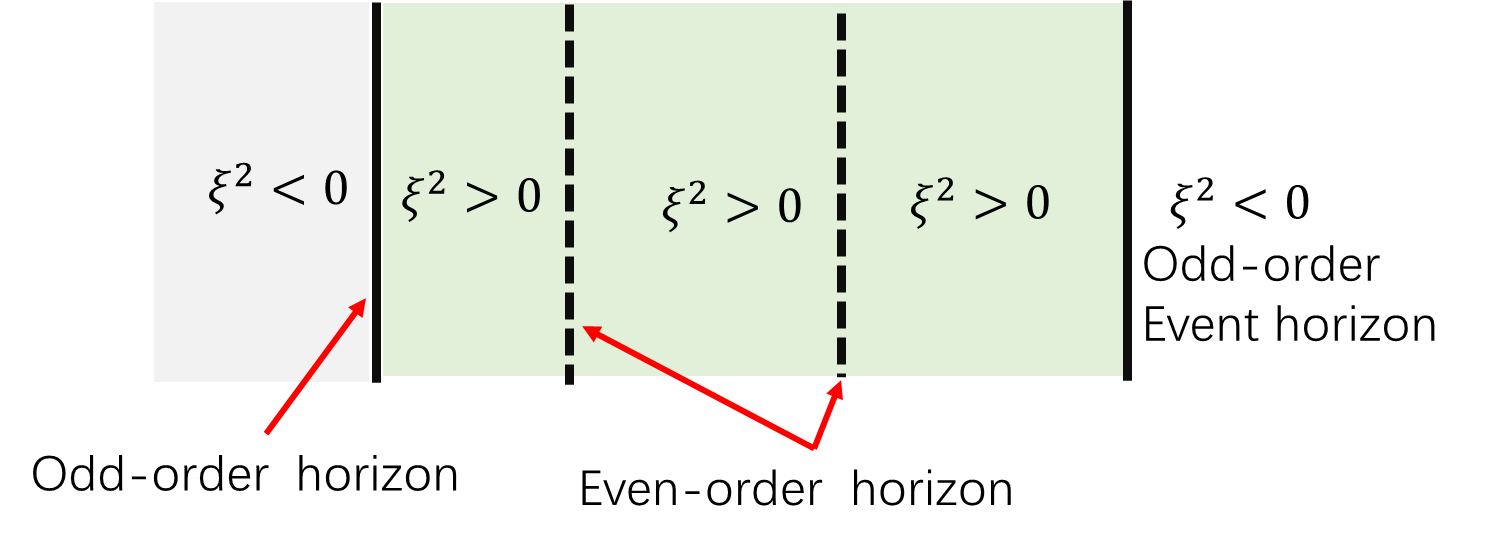}
   \caption{Our results do not restrict the number of even-order horizons between an odd-order event horizon and its odd-order inner horizon.  }\label{fighs2a}
\end{figure}

In \textbf{Theorem 1}, we have restricted ourselves to be the case for which all horizons have non-zero surface gravity. Nevertheless, our discussion can be generalized into the case with degenerate horizons, although degenerate horizons are not generally considered to be of direct physical significance. We first notice that, if the surface gravity of a horizon is zero, then $\xi^2$ can have different or same sign at two sides of the horizon. To distinguish such two different cases, we introduce the conception ``even/odd order horizon'' : the horizon with different/same sign of $\xi^2$ at its two sides is called an odd-order/even-order. By this definition, we can conclude that: (1) if there are two odd-order inner horizons inside an odd-order event horizon, or (2) if there is an inner horizon inside an even-order horizon, then there is spacetime $\mathcal{V}$ which is bounded by two null surfaces of $\xi^2=0$ (see Fig.~\ref{fighs1b} for illustration). Following our proof for the non-degenerate case, we can obtain the following result:
\begin{theorem}
If any one of the conditions C1-C3 in \textbf{Theorem 1} is satisfied, then there is at most one odd-order inner Killing horizon associated with $\xi^\mu$ inside every connected branch of  event horizon. In addition, if a connected branch of black hole event horizon is even-order, then there is no inner Killing horizon associated with $\xi^\mu$.
\end{theorem}
Note that non-degenerate horizon is an odd-order horizon by our definition. Therefore, \textbf{Theorem 2} includes \textbf{Theorem 1} as its one special case. We should mention that, if the event horizon is odd-order, our results do not restrict the number of even-order horizons between the event horizon and its odd-order inner horizon (see Fig.~\ref{fighs2a} for a schematic explanation).

Our results are applicable to the asymptotically flat and AdS cases, similar arguments and methods could also be generalized to the asymptotically dS spacetime case. One significant difference in the dS case is that there is a positive cosmological constant which itself violates the SEC. As a consequence, outside the event horizon there is a ``cosmological horizon'' beyond which the Killing vector $\xi^\mu$ becomes spacelike. Nevertheless, following the similar steps one can find that: if one of C1-C3 is satisfied inside the cosmological horizon, then (i) there is no horizon inside every connected branch of cosmological horizon with non-zero surface gravity and (ii) there is at most one horizon inside every connected branch of even-order degenerate cosmological horizon. In the asymptotically dS case, the cosmological constant always dominates and violates SEC at infinity. However, our condition C1 only requires SEC to be satisfied inside the cosmological horizon.  For the case with a compact horizon, if classical matters are strong enough inside the cosmological horizon, it can prevent the formation of inner horizons. A trivial example is the RN-dS black hole. Besides the cosmological horizon, there is an event horizon as well as an inner horizon behind the event horizon when the U(1) charge is small. However, when the charge is large enough to dominate the interior behind the cosmological horizon so that the SEC~\eqref{stronge1} is satisfied, there is no horizon except the cosmological horizon.

\begin{acknowledgments}
This work was partially supported by the National Natural Science Foundation of China Grants No.12122513, No.12075298, No.11821505, No.11991052, No.12047503, No.11690022 and No.12005155, and by the Key Research Program of the Chinese Academy of Sciences (CAS) Grant NO. XDPB15, the CAS Project for Young Scientists in Basic Research YSBR-006 and the Key Research Program of Frontier Sciences of CAS.
\end{acknowledgments}

\appendix
\section{Horizons of RN black hole}\label{rncase}
In this section, we illustrate a possible issue for counting the number of horizon in terms of coordinate gauge. In contrast, such ambiguity can be definitely avoided with our definition of the  back hole interior.
As an illustrative example, we consider the 3+1 dimensional RN black hole for which as is well accepted in the literature that it has two horizons, including an event horizon and a Cauchy horizon inside.  The metric for the RN black hole in the standard coordinates reads
\begin{equation}\label{metricrn1}
  \mathrm{d}s^2=-f(r)\td t^2+f(r)^{-1}\td r^2+r^2(\td\theta^2+\sin^2\theta\td\varphi^2)\,,
\end{equation}
with the blackening function $f(r)=(r-r_H)(r-r_-)/r^2$. It is manifest that besides a time-like singularity at $r\rightarrow 0$, it has an event horizon at $r_H$ associated with the Killing vector $\xi^\mu=(\frac{\partial}{\partial t})^\mu$  and an inner Cauchy horizon at $r_-$. It is clear that $\xi^2=-f(r)$ has two roots $r_H$ and $r_-$, thus one may conclude that there are two horizons. However, Eq.~\eqref{metricrn1} is not the only static spherically symmetric coordinate gauge. Let us define $\rho^2=r-r_H$. The coordinates $\{t,\rho,\theta,\varphi\}$ can still keep $t$ to be the orbit of time-like Killing vector and $\{\theta,\varphi\}$ to present the spherical symmetry. In this coordinate gauge the metric becomes
\begin{equation}\label{metricrn2}
  \mathrm{d}s^2=-\frac{(r_H-r_-+\rho^2)\rho^2}{(r_H+\rho^2)^2}\td t^2+\frac{4(\rho^2+r_H)^2}{r_H-r_-+\rho^2}\td\rho^2+(r_H+\rho^2)(\td\theta^2+\sin^2\theta\td\varphi^2)\,.
\end{equation}
One then immediately obtains that
%
\begin{equation}\label{newxi2}
  \xi^2=-\frac{(r_H-r_-+\rho^2)\rho^2}{(r_H+\rho^2)^2}\,,
\end{equation}
which has only one real root at $\rho=0$. Thus we may conclude that the RN black hole has only one horizon if we use the new coordinates. One could argue that the coordinate gauge~\eqref{metricrn2} can only cover some parts of the RN spacetime and the coordinate gauge~\eqref{metricrn1} can cover regions larger than~\eqref{metricrn2}. However, it needs to note that the coordinate gauge~\eqref{metricrn1} can be analytically extended. If we use  maximally analytical continued RN spacetime, we will find that the number of horizons becomes infinite (see Fig.~\ref{figrn1} for illustration). According to our definition for the interior of back holes introduced in the main text, it is manifest that the RN black hole has two horizons, agreeing  with the usual statement  in the literature.
%


\section{Black holes with maximal symmetric horizons}\label{app1}
We give explicitly the expressions on $F(z)$ and $I(z)$ for black holes with planar, hyperbolically and spherically symmetric  horizon cases.
We foliate the spacetime by
\begin{equation}\label{metric1}
  \mathrm{d}s^2=z^{-2}[-f(z)e^{-\chi(z)}\td t^2+f(z)^{-1}\td z^2+\td\Sigma_{k,d-1}^2]\,,
\end{equation}
where $\td\Sigma_{k,d-1}^2$ is the standard metric of planar ($k=0$), unit sphere ($k=1$) or unit hyperbolic plane ($k=-1$). Then we have $N^2=z^{-2}f(z)e^{-\chi(z)}, \varphi^2=1/(z^2f(z)), N_A=0$ and $q_{AB}\td x^A\td x^B=z^{-2}\td\Sigma_{k,d-1}^2$.
The normal vector of $\mathcal{S}_z$ in all cases is
\begin{equation}\label{binormalv1}
  s^a=z\sqrt{f}(\partial/\partial z)^a\,,
\end{equation}
and we have
\begin{equation}\label{dxieq2}
  s^a\partial_aN=z\sqrt{f}\partial_z(\sqrt{fe^{-\chi}/z^2})=\frac{e^{\chi/2}}2z^2\partial_z(fe^{-\chi}/z^2)=\frac12[e^{\chi/2}(fe^{-\chi})'-2fe^{-\chi/2}/z]\,.
\end{equation}
The extrinsic curvature of $\mathcal{S}_z$ reads
\begin{equation}\label{extrink1}
  \mathcal{K}_{AB}=\frac{\sqrt{f}z}2\partial_zq_{AB}\,,
\end{equation}
where $q_{AB}$ is the induced metric of $\mathcal{S}_z$. For the metric (\ref{metric1}), we have $\partial_zq_{AB}=-2q_{AB}/z$ and $\mathcal{K}=-(d-1)\sqrt{f}$, which yields
\begin{equation}\label{extrink2}
  \mathcal{K}N=-(d-1)\sqrt{f}\frac{\sqrt{f}e^{-\chi/2}}z=-(d-1)fe^{-\chi/2}/z\,.
\end{equation}
Thus, we find
\begin{equation}\label{defQT4}
  F(z)=\frac12\mathcal{A}^{-1}\int_{\mathcal{S}_{z}}e^{\chi/2}(fe^{-\chi})'\td S\,.
\end{equation}
Here, $\mathcal{A}=z^{d-1}\int\td S=\int\td\Sigma_{k,d-1}$ is independent of $z$, so we have
\begin{equation}\label{defQT4}
  F(z)=\frac12z^{1-d}e^{\chi/2}(fe^{-\chi})'\,.
\end{equation}
The energy momentum tensor takes the diagonal form ${\hat{T}^\mu}_{~~\nu}$=diag$(-\hat{\rho}, \hat{p}_z, \hat{p}, \cdots,\hat{p})$. Then we then have  $\varpi=d\hat{\rho}+\hat{T}-\hat{p}_z=(d-1)(\hat{\rho}+\hat{p})$ and
\begin{equation}\label{nulkpanar2}
\begin{split}
  I(z)=&\frac1{(d-1)\mathcal{A}}\int\left[N\varphi(\varpi-\mathfrak{R})\right]\td S\\
  =&\frac1{(d-1)\mathcal{A}}\int z^{-2}e^{-\chi/2}(d-1)[\hat{\rho}+\hat{p}-(d-2)kz^2]z^{1-d}\td\Sigma_{k,d-1}\\
  =&z^{-1-d}e^{-\chi/2}[\hat{\rho}+\hat{p}-(d-2)kz^2]\,.
  \end{split}
\end{equation}
Thus, Eq.~\eqref{fizeq1} implies that $F(z)-\int^zI(x)\td x$ is independent of $z$. As a double check, we have verified this result by using the Einstein's equation under the metric ansatz~\eqref{metric1}.

Eq.~\eqref{surint4} in the main text is a purely mathematical identity. As a double check, we compute the right hand side of Eq.~\eqref{surint4} under the metric~\eqref{metric1}. Let us define the right hand side of Eq.~\eqref{surint4} to be $\mathcal{I}(z)$. Under~\eqref{metric1}, it reduces to
\begin{equation}\label{surint4b}
\mathcal{I}(z):=\frac1{(d-1)}[(d-1)\varphi D^2N+{^{(d)}}RN\varphi/2-N\varphi G_{ab}s^as^b-N\varphi\mathfrak{R}]z^{1-d}\,.
\end{equation}
We first compute the term $G_{ab}s^as^b=G_{\mu\nu}s^\mu s^\nu$, which yields
\begin{equation}\label{Gzz1}
  \varphi N G_{ab}s^as^b=fe^{-\chi/2}G_{zz}=\frac{d-1}{2z^2}[df-zf'-(d-2)kz^2+fz\chi']e^{-\chi/2}\,.
\end{equation}
Straightforward computation gives
\begin{equation}\label{d2N1}
  \varphi D^2N=z^{d-1}\partial_z\left[z^{2-d}\sqrt{f}(\sqrt{f}e^{-\chi/2}/z)'\right]\,,
\end{equation}
\begin{equation}\label{d2N2}
  {^{(d)}}RN\varphi=(d-1)[zf'+(d-2) kz^2-df]z^{-2}e^{-\chi/2}\,,
\end{equation}
and
\begin{equation}\label{eqfor2R}
  \mathfrak{R}=(d-1)(d-2)kz^2\,.
\end{equation}
Then it is easy to check that $F'(z)=\mathcal{I}(z)$ is true for arbitrary $f(z)$ and $\chi(z)$.

%


%

\end{document}